\documentclass[twocolumn,
    preprintnumbers,
    amsmath,
    amssymb,
    nofootinbib,
    longbibliography,
    tightenlines,
    nobibnotes,
    prb]{revtex4-1}
\usepackage{amsfonts}
\usepackage[english]{babel}
\usepackage[T1]{fontenc}
\usepackage{times}
\usepackage{mathrsfs}
\usepackage{graphicx}
\usepackage{dcolumn}
\usepackage{bm}
\usepackage[colorlinks,bookmarks=true,citecolor=blue,linkcolor=red,urlcolor=blue]{hyperref}
\usepackage[tight, FIGTOPCAP, hang, raggedright, nooneline]{subfigure}

\DeclareMathOperator{\Tr}{Tr}

\renewcommand{\vec}[1]{\boldsymbol{\mathbf{#1}}}
\newcommand{\vk}{\vec k}
\newcommand{\vq}{\vec q}

\newcommand{\vsigma}{\mbox{\boldmath $\sigma$}}

\usepackage{xcolor}

\begin{document}

\title{Tilted disordered Weyl semimetals}
\author{Maximilian Trescher,$^1$ Bj\"orn Sbierski,$^1$ Piet W. Brouwer,$^1$ and Emil J. Bergholtz$^{1,2}$ }
\affiliation{$^1$ Dahlem Center for Complex Quantum Systems and Institut f\"ur Theoretische Physik, Freie Universit\"at Berlin, Arnimallee 14, 14195 Berlin, Germany
\\ $^2$ Department of Physics, Stockholm University, AlbaNova University Center, 106 91 Stockholm, Sweden} 
\date{\today}

\begin{abstract} 
Although Lorentz invariance forbids the presence of a term that ``tilts'' the energy-momentum relation in the Weyl Hamiltonian, a tilted dispersion is not forbidden and, in fact, generic for condensed matter realizations of Weyl semimetals. We here investigate the combined effect of such a tilted Weyl dispersion and the presence of potential disorder. In particular, we address the influence of a tilt on the disorder-induced phase transition between a quasi-ballistic phase at weak disorder, in which the disorder is an irrelevant perturbation, and a diffusive phase at strong disorder. Our main result is that the presence of a tilt leads to a reduction of the critical disorder strength for this transition or, equivalently, that increasing the tilt at fixed disorder strength drives the system through the phase transition to the diffusive strong-disorder phase. Notably this obscures the tilt induced Lifshitz transition to an over tilted type-II Weyl phase at any finite disorder strength. Our results are supported by analytical calculations using the self-consistent Born approximation and numerical calculations of the density of states and of transport properties.
\end{abstract}

\maketitle

\section{Introduction}

The theory of Weyl semimetals---semimetals where non-degenerate conduction and valence bands have touching points---has a long and intriguing history, bringing ideas originally developed in the context of particle physics into the realm of condensed matter and materials physics.\cite{weyl_elektron_1929,volovik_universe_2009, murakami_phase_2007,wan_topological_2011,burkov_weyl_2011} Recently the experimental discovery of a Weyl semimetal was reported by various groups,\cite{xu_discovery-taas_2015,lv_discovery_2015,lu_experimental_2015} soon followed by transport measurements demonstrating the chiral anomaly.\cite{huang_observation_2015,zhang_signatures_2016}

\begin{figure}[ht]
    \centering
    \includegraphics[width=0.23\textwidth]{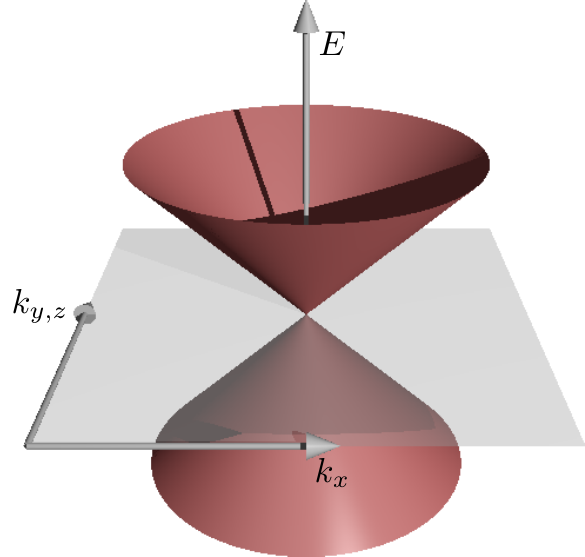}
    \includegraphics[width=0.23\textwidth]{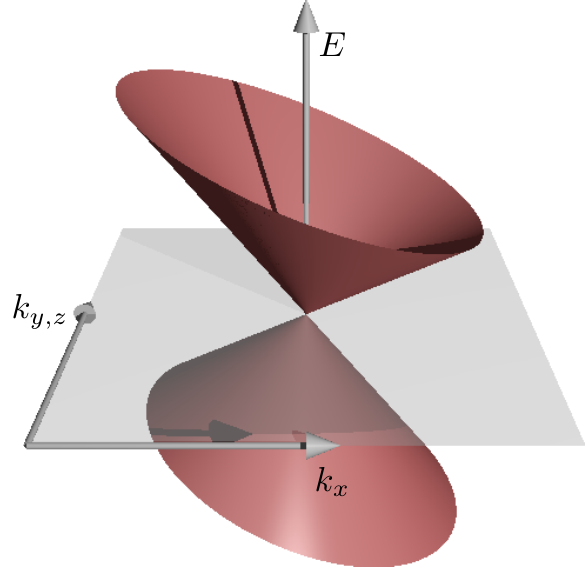}
    \caption{(Color online) Illustration of a tilted Weyl dispersion, projected to two momentum dimensions. The left panel shows the isotropic case ($a=0$); the right panel shows a dispersion tilted in the negative $x$ direction with tilt parameter $a = 0.4$.}
\label{fig:illustration}
\end{figure}

The conventional understanding of a Weyl semimetal is a system that, in the vicinity of the band touching points, is accurately described by the two-band Weyl Hamiltonian
\begin{align}
  H_{\textrm{0}} &= \hbar v \vec{k}\cdot \vsigma,
  \label{H}
\end{align}  
whose elegant form is dictated by Lorentz invariance and the requirement of a linear dispersion of the crossing bands. Here, momentum and energy are measured relative to the momentum and energy of the band crossing point, and $\vsigma = (\sigma_1,\sigma_2,\sigma_3)$ is the Pauli matrix. However, as was only appreciated recently, symmetry breaking terms such as an anisotropic velocity and a ``tilt'' of the Weyl dispersion (see Fig.~\ref{fig:illustration}) occur generically in Weyl materials and may have profound consequences for thermodynamic and transport properties.\cite{bergholtz_topology_2015,trescher_quantum_2015,soluyanov_type-ii_2015,
xu_anisotropic_2014,
xu_structured_2015,rodionov_effects_2015,jccm_2015_beenakker,
xu_type2-coldatoms_2016,
deng_experimental_2016,xu_discovery-weyl2-mote2_2016,huang_spectroscopic_2016,wang_spectroscopic-weyl2-wte2_2016,belopolski_measuring-weyl2-mowte2_2016,koepernik_tairte4-ternary-weyl2_2016} Explicitly, a generic linear band crossing is described by
\begin{align}
    H_{\textrm{0}} &= \sum_{i,j=1}^{3} \hbar v_{ij} k_i \sigma_j + 
    \sum_{i=1}^{3} v_{ii} a_i k_i \sigma_0 \ , \label{general_H}
\end{align}
where the $v_{ij}$ now describe an anisotropic velocity, the $a_i$ represent a uniform linear tilting of both bands, and $\sigma_0$ is the $2 \times 2$ unit matrix. While the tilt looks most innocent---being proportional to the unit matrix $\sigma_0$ it does not alter the eigenstates of the model---it turns out to be the more interesting symmetry-breaking term. Whereas tilting terms have been considered early on in two-dimensional systems as consequences of perturbations,\cite{2d_tilt,2d_tilt2} the topological stability of Weyl points,\cite{wan_topological_2011} or, viewed alternatively, the generic nature of non-degenerate band crossings\cite{herring} in three dimensions allows for much stronger tilting. In fact, it was first shown in simple toy models,\cite{bergholtz_topology_2015} and later in more realistic materials simulations\cite{soluyanov_type-ii_2015} and in models of superfluids,\cite{xu_structured_2015} that the Weyl dispersion can easily be tilted over to the extent that a finite Fermi surface is formed, with hole and electron pockets touching at the Weyl point. These systems have been called ``Type-II Weyl semimetals'' and have, following their prediction in WTe$_2$, now been observed in a number of materials.\cite{deng_experimental_2016,xu_discovery-weyl2-mote2_2016,huang_spectroscopic_2016,wang_spectroscopic-weyl2-wte2_2016,belopolski_measuring-weyl2-mowte2_2016,koepernik_tairte4-ternary-weyl2_2016}

The effect of smaller, sub-critical tilts, which preserve the point-like nature of the Fermi surface, is more of a quantitative than of a qualitative nature, although even small tilts influence transport properties of Weyl semimetals in a unique way. When the Fermi level lies exactly at the nodal point, a Weyl semimetal of size $L^3$ has a vanishing conductivity $\sigma$, but a non-vanishing $L$-independent conductance $g = \sigma L$.\cite{baireuther_quantum_2014,sbierski_quantum_2014} The finite conductance is reminiscent of the universal minimal conductance $g = e^2/4 \pi h$ predicted and observed in graphene,\cite{tworzydlo_sub-poissonian_2006,katsnelson_zitterbewegung_2006} with the important distinction that in two dimensions a finite minimal conductance corresponds to a non-zero conductivity due to its different scaling with system size ($g = \sigma L^{d-2}$ in $d$ dimensions). As shown in Ref.\ \onlinecite{trescher_quantum_2015}, the value of the minimal conductance $g$ in a Weyl semimetal depends on the tilt. Moreover, tilt was also found to affect the value of the ``Fano factor'' $F = P/2 e I$ at the Weyl point (the ratio of shot noise power $P$ to current $I$ \cite{trescher_quantum_2015}), which was otherwise found to be a universal number $F = (1 + 2 \ln 2)/(6 \ln 2)$,\cite{baireuther_quantum_2014,sbierski_quantum_2014} independent of weak disorder, sample geometry, or anisotropy in the Weyl dispersion.

In this work we consider the combined effect of a (sub-critical) tilt of the Weyl dispersion and the presence of potential disorder, which is the second important addition to the Weyl Hamiltonian (\ref{H}) required for the description of realistic condensed-matter realizations of a Weyl semimetal. For isotropic Weyl cones the remarkable properties of the ideal system---in addition to the anomalous transport properties described above these include the vanishing density of states at the Weyl point---are known to persist up to a finite disorder strength, beyond which a diffusive phase with nonzero density of states at the Weyl point and nonzero conductivity sets in.\cite{fradkin_critical_1986,sbierski_quantum_2014,sbierski_2016,kobayashi_2014,syzranov_2015,sbierski_quantum_2015,pixley_2015,louvet_2016,roy_2016,syzranov_2016} The Fano factor $F$ takes a different, but again universal value $F = 1/3$ in the diffusive phase.

Our main finding, which we support with a combination of analytical and numerical arguments, is that the main effect of tilting the Weyl dispersion is to reduce the critical disorder strength $K_{\rm c}$ for the transition between the ``pseudo-ballistic'' weak-disorder phase and the diffusive strong-disorder phase. Such a result can be expected intuitively, based on the consideration that in the weak-disorder phase there is a vanishing density of states at the Weyl point, and disorder acts by virtual excitations to states of higher energy, where the range of reachable energies is determined by the disorder strength.\cite{syzranov_2015,syzranov_2016} Tilting the Weyl dispersion lowers the energy of some states while raising the energy of others. The net effect, however, is that there are more states available in a given energy range than in the isotropic cone, see Fig.~\ref{fig:illustration}. Explicitly, the density of states $\nu(\varepsilon)$ depends on the tilt parameter $a$ as
\begin{align}
    \nu(\varepsilon) &=
    \frac{\nu_0(\varepsilon)}{\left( 1-a^2 \right)^2}
 \geq \nu_0(\varepsilon ) , 
    \label{dos-tilt}
\end{align} 
where $\nu_0(\varepsilon)$ is the density of states without tilt. This implies that the disorder of the same strength $K$ can be expected to have a stronger influence on a tilted Weyl dispersion, and it is natural to expect a lower critical disorder strength $K_{c}$ for cones with larger tilt. 

Upon approaching the critical tilt $a=1$ the critical disorder strength $K_{\rm c}$ goes to zero---again, not a surprise given the diverging density of states for $a \to 1$. This implies, however, that in the presence of disorder the tilt-driven transition between type-I and type-II Weyl semimetals will be preceded (and, hence, masked) by the tilt-driven transition from the pseudo-ballistic weak-disorder phase into the diffusive strong-disorder phase.

The remainder of this work is organized as follows: In Sec.\ \ref{sec:setup} we introduce to the specific model employed throughout this work, which contains a tilted dispersion as well as a random potential. In Sec.\ \ref{sec:scba} we employ the self-consistent Born approximation to show how a tilted dispersion affects the disorder-induced phase transition. We then corroborate these results by two distinct numerical approaches: We use the kernel polynomial method\cite{weisse_kpm_2006} to compute the density of states in a tight-binding model using the same disorder type as in the SCBA, see Sec.~\ref{sec:kpm}, and we study transport properties in finite size systems using a scattering matrix approach, see Sec.\ \ref{sec:transport}. The use of numerical methods to confirm our conclusions is necessary, since the self-consistent Born approximation is known to be an uncontrolled approximation for the disorder-induced phase transition in Weyl semimetals.\cite{sbierski_quantum_2014} We conclude in Sec.\ \ref{sec:discussion}.

\section{Model} 
\label{sec:setup}

We consider a single disordered Weyl node described by the Hamiltonian
\begin{align}
  H = H_0 + U,\ \ H_{0} = 
  v (\vec{k} \cdot \vsigma + a k_z \sigma_0), \label{lowEH}
\end{align}
where without loss of generality we have chosen the tilt to be in the $z$ direction. Momentum and energy are measured with respect to the Weyl node. We have chosen units such that $\hbar = 1$. The critical tilt is $a=1$, and we will consider sub-critical tilts $0 \le a < 1$ only.

The disorder potential $U$ is taken to be a Gaussian random potential with zero average and with two-point correlation function
\begin{align}
    \langle U(\vec{q}) U(\vec{q}')\rangle &= \frac{K \xi h^2 v^2}{L^3} e^{-\frac{q^2 \xi^2}{2}} \delta_{\vq,\vq'} 
    \label{correlator-finite-size}
\end{align}
for a finite system of size $L^3$. Here $\xi$ is the correlation length of the disorder potential and $K$ is the dimensionless disorder strength. The restriction to a single Weyl node requires that the correlation length $\xi$ is much larger than the inverse distance between the Weyl nodes. 

We have chosen not to include anisotropy of the Weyl dispersion in to the Hamiltonian (\ref{lowEH}) [compare with Eq.\ (\ref{general_H})]. Although anisotropies are as ubiquitous in real materials as the tilts, they can largely be understood by simple means of rescaling of the coordinate axes, rendering the Weyl dispersion isotropic but the disorder anisotropic.\cite{trescher_quantum_2015,rodionov_effects_2015} Through such a rescaling procedure, anisotropies were found to affect the conductance $g$ (since $g$ depends on sample geometry), but not the Fano factor $F$. Tilts, on the other hand, can not be removed by the rescaling procedure, and were found to affect both $g$ and $F$ in the absence of disorder.\cite{trescher_quantum_2015} 

In the following three Sections we calculate the critical disorder strength $K_{\rm c}$ for the disorder-induced transition to a diffusive phase using three different methods. Section \ref{sec:scba} employs the self-consistent Born approximation and addresses the density of states, using the model described above. Section \ref{sec:kpm} uses the kernel-polynomial method to calculate the density of states for a tight-binding model for which the continuum model (\ref{lowEH}) is the low-energy limit. Section \ref{sec:transport} considers transport properties of the continuum model (\ref{lowEH}), but for a system of size $W^2 \times L$, where the sample width $W$ (transverse to the transport direction) is chosen much larger than its length $L$ (in the transport direction) to ensure that the transport properties do not depend on the choice of the boundary conditions in the direction transverse to the current flow.

\section{Density of states from SCBA}

\label{sec:scba}

To study the density of states of the model (\ref{lowEH}) we employ the self-consistent Born approximation (SCBA). We closely follow a similar calculation of Ominato and Koshino for a Weyl dispersion without tilt.\cite{ominato_quantum_2014} 

Before we turn to a description of our calculations, we note that the SCBA relies on a diagrammatic expansion that is known to neglect important contributions. For a Weyl dispersion without tilt, the SCBA is known to yield critical disorder strengths $K_c$, which are larger by a factor $\sim 2$ than the ones obtained by more precise numerical simulations.\cite{sbierski_quantum_2014} We nevertheless see the SCBA as a useful approximation for a qualitative understanding of the way in which a tilted Weyl dispersion affects the critical disorder strength. Furthermore, the SCBA allows us to access the disorder-induced renormalizations of tilt and Fermi velocity.

\begin{figure}[ht]
    \includegraphics[width=0.48\textwidth]{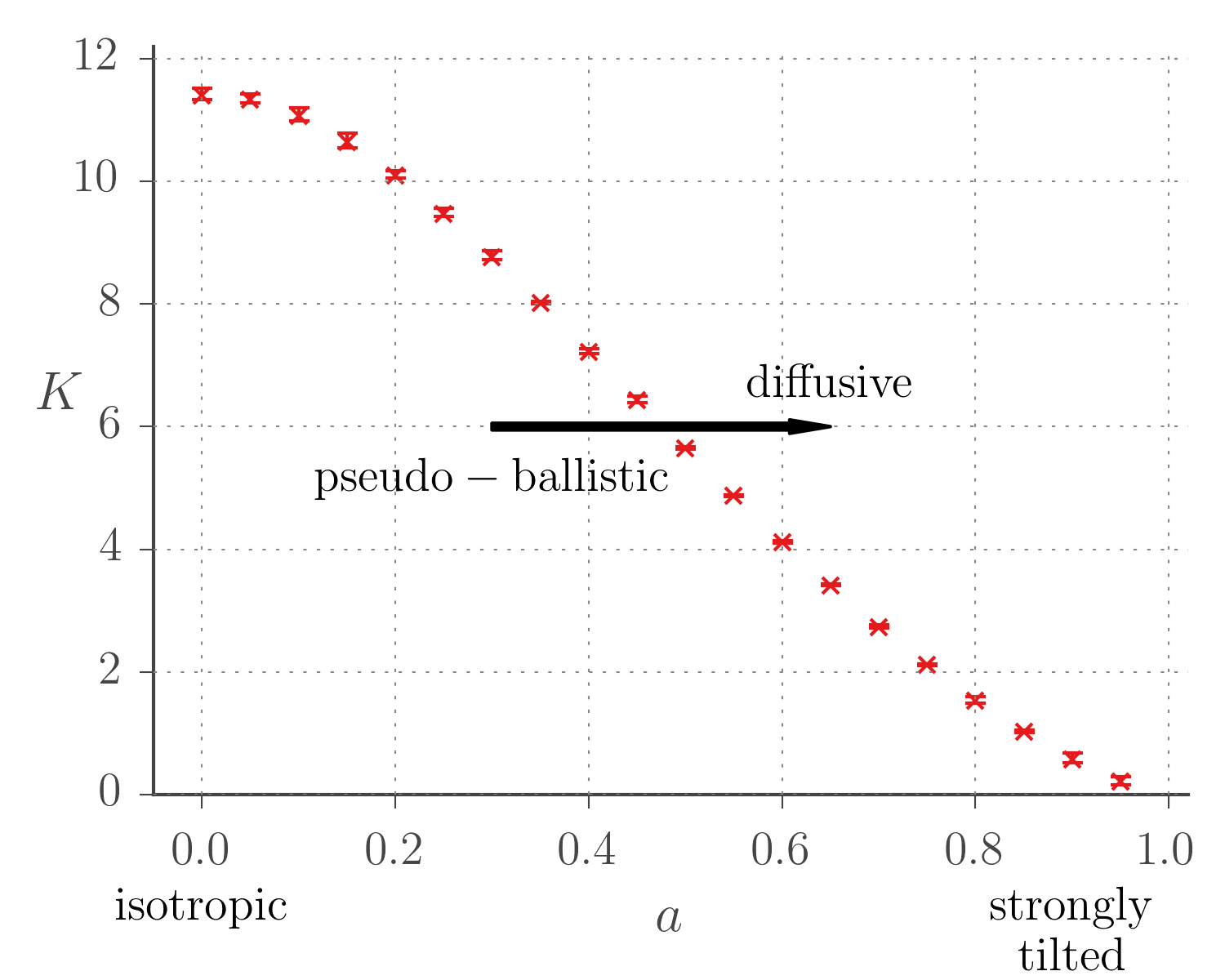}
    \caption{(Color online) Phase diagram for the tilted Weyl cone obtained from the SCBA. The numerical uncertainty for $K_{\rm c}$ is shown as errorbars (see appendix for a discussion). 
}
    \label{fig:weyl-tilted}
\end{figure}

The disorder-averaged density of states $\nu(\varepsilon)$ is expressed in terms of the ($2 \times 2$ matrix) Green function as
\begin{align}
   \nu(\varepsilon) = - \frac{1}{\pi L^3} \textrm{Im} \sum_{\vk} \Tr \left\langle G(\vec{k}, \varepsilon + i0^+) \right\rangle,
    \label{dos}
\end{align}
where the brackets $\langle \ldots \rangle$ indicate the disorder average. In the SCBA the disorder-averaged Green function is expressed as
\begin{align}
    \langle {G}(\vec{k}, \varepsilon) \rangle &= 
    \frac{1}{(\varepsilon - v a k_z) \sigma_0 - v \vec{k}\cdot\vec{\sigma} - \Sigma(\vec{k}, \varepsilon)},
    \label{greens}
\end{align}
with the SCBA approximation for the self-energy 
\begin{align}
    \Sigma(\vec{k}, \varepsilon) 
    &= \sum_{\vk'} G(\vec{k}', \varepsilon)  \langle |U(\vec{k}-\vec{k'})|^2\rangle,
    \label{selfenergy}
\end{align}
where the disorder average $\langle |U(\vq)|^2 \rangle$ is given by \eqref{correlator-finite-size}. The self-consistent integral equations \eqref{greens} and \eqref{selfenergy} are solved numerically. Details on the numerical procedure are given in appendix \ref{scba-appendix}. The critical disorder strength $K_{\rm c}$ is found as that disorder strength for which the density of states $\nu(0)$ at the Weyl point becomes finite, whereas $\nu(0) = 0$ for $K < K_{\rm c}$. Our main results are summarized in Fig.~\ref{fig:weyl-tilted}.

The SCBA phase diagram of Fig.\ \ref{fig:weyl-tilted} confirms the intuitive picture of the introduction: the critical disorder strength decreases with increasing tilt and approaches $0$ when the tilt strength $a$ approaches the critical value $a=1$. This is consistent, since as a finite density of states develops when the cone tips over ($a \geq 1$), we expect a diffusive phase at arbitrarily small disorder strengths for super-critical tilt.

\begin{figure}[hb]
    \centering
    \includegraphics[width=0.5\textwidth]{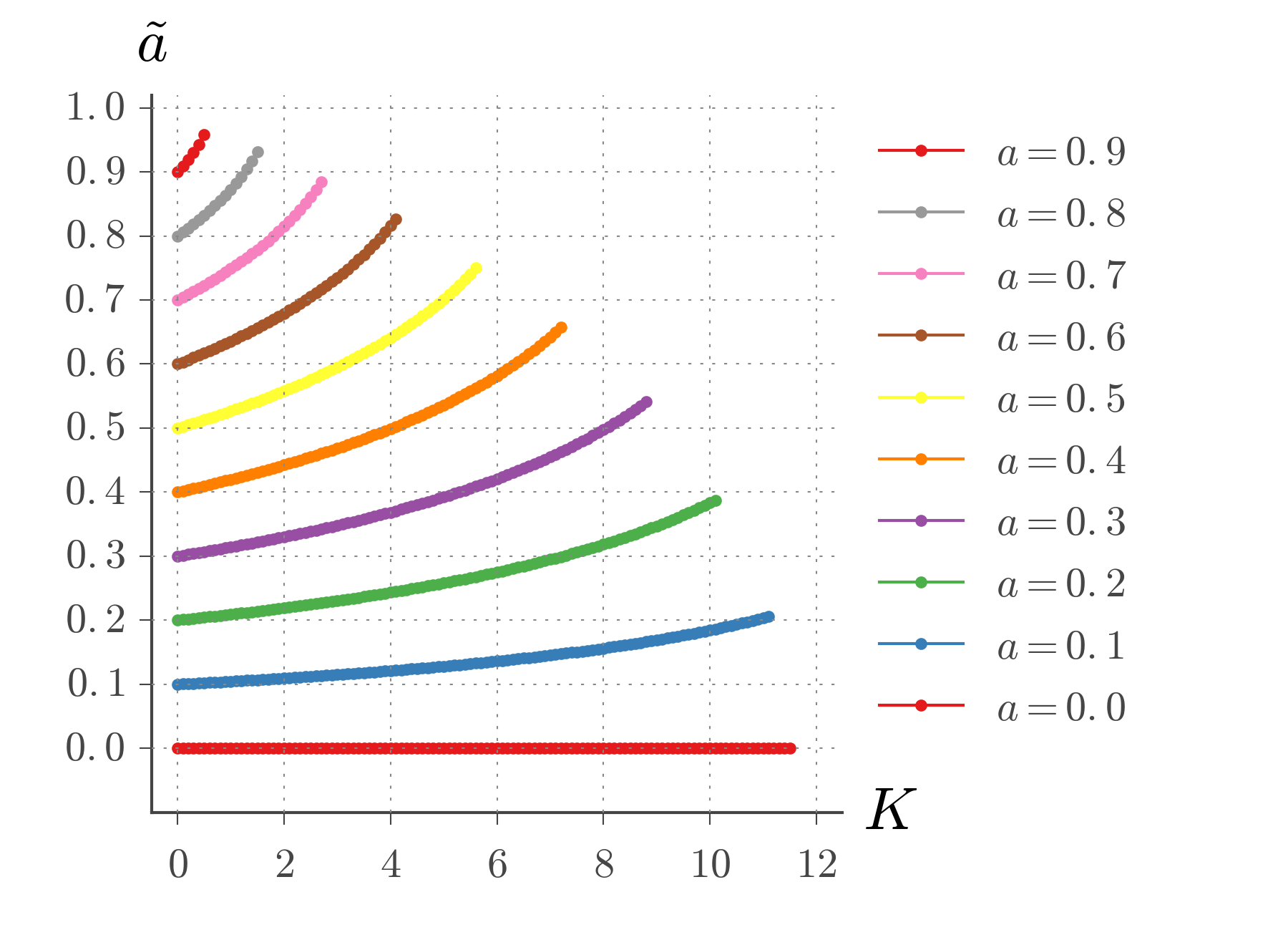}
    \caption{Effective SCBA tilt $\tilde a$ vs.\ dimensionless disorder strength $K$ in the weak-disorder phase $K < K_{\rm c}$.}
    \label{fig:effective-tilt}
\end{figure}

The SCBA not only allows us to find the tilt-dependence of the critical disorder strength $K_{\rm c}$, the expansion of the self energy $\Sigma(\vk,0)$ for small momenta at zero energy also allows us to find a disorder-renormalized Fermi velocity $\tilde v$ and tilt $\tilde a$. Taking into account rotation invariance around the $z$ axis, the small-$\vk$ expansion of the self energy reads 
\begin{equation}
  \Sigma(\vk,0) = v (\alpha \vk \cdot \sigma + \beta a k_z \sigma_0)
  + {\cal O}(k^2),
\end{equation}
from which we obtain
\begin{equation}
  \tilde v = v (1 + \alpha),\ \
  \tilde a = a \frac{1 + \beta}{1 + \alpha}.
  \label{effective-parameters}
\end{equation}

Our result for the renormalized Fermi energy $\tilde v$ is consistent with that of Ref. \onlinecite{ominato_quantum_2014} and will not be discussed further here. The renormalized dimensionless tilt $\tilde a$ is shown in Fig.~\ref{fig:effective-tilt}, as a function of disorder strength $K$. 
Figure \ref{fig:effective-tilt} suggests that there is not only the single fixed point of the clean and isotropic Weyl cone (in the renormalization-group sense), but rather a continuous family of fixed points at no disorder which are distinguished by their tilt. Also note that while disorder $K < K_{\rm c}$ leads to an increased dimensionless tilt, in the SCBA approximation sub-critical disorder can not lead to an ``over-tilting'' of the Weyl dispersion. In other words, the disorder-induced phase transition to a diffusive phase with a finite density of states at the Fermi level always takes place {\em before} the tilt-induced transition to a type-II Weyl semimetal, which also has a finite density of states at the nodal point. (This is consistent with the tilt-dependence of the critical disorder strength shown in Fig.\ \ref{fig:weyl-tilted}, which shows that the critical disorder strength approaches zero in the limit $a \to 1$.) This observation remains valid if the disorder-induced renormalization of the tilt is included, see Fig.\ \ref{fig:effective-tilt}.

\section{Density of states from KPM}
\label{sec:kpm}

To complement the SCBA analysis we now report a numerical calculation of the density of states, using the kernel polynomial method (KPM).\cite{weisse_kpm_2006}
The KPM is a numerically efficient method to approximate the density of states of large lattice Hamiltonians $H$ represented as sparse matrices. As a first step, the Hamiltonian is rescaled so that the spectrum fits in the interval $[-1,1]$. The density of states is then expanded in Chebyshev polynomials. The expansion coefficients (up to a certain order, usually a few thousand) can be expressed as a trace over a polynomial of $H$ which can be well approximated numerically by using a few random states. Employing identities for Chebyshev polynomials, the expansion coefficients can be efficiently calculated by iteration involving only matrix-vector products. Finally, residual Gibbs oscillations in the density of states are suppressed using an appropriate smoothing Kernel.

We consider the two-band lattice model\cite{turner_beyond_2013}
\begin{align}
  H_{0}\left(\mathbf{k}\right) =&
  \frac{v}{b} \left[\sigma_{x}\sin bk_{x}+\sigma_{y}\sin bk_{y} \right. \nonumber \\
& \left. -\sigma_{z}\cos bk_{z}-\sigma_0 a\cos bk_{z}\right] \label{eq:H0}
\end{align}
where $b$ is the lattice constant. The model (\ref{eq:H0}) features eight Weyl points at momenta $\mathbf{k}(\tau_x,\tau_y,\tau_z) = (k_x(\tau_x),k_y(\tau_y),k_z(\tau_z))$ for $\tau_{x,y,z} = \pm 1$ with $k_x(\tau_x) = (\pi/2 b) (1 - \tau_x)$, $k_{y}(\tau_y)=(\pi/2 b)(1 - \tau_y)$, $k_{z}(\tau_z)= \tau_z \pi/2b$. We add a disorder
potential as in Eq. \eqref{correlator-finite-size}, with correlation length $\xi =5b$, which ensures that (i) the smooth disorder correlations are well represented on the discrete
lattice and (ii) the inter-node scattering rate is suppressed as compared to the intra-node rate by a factor of order $\exp[-\pi^{2}\xi^{2}/2 b^2]<10^{-19}$, such that the physics can be regarded as effectively single-node.

\begin{figure}[h]
\begin{centering}
    \includegraphics[width=0.48\textwidth]{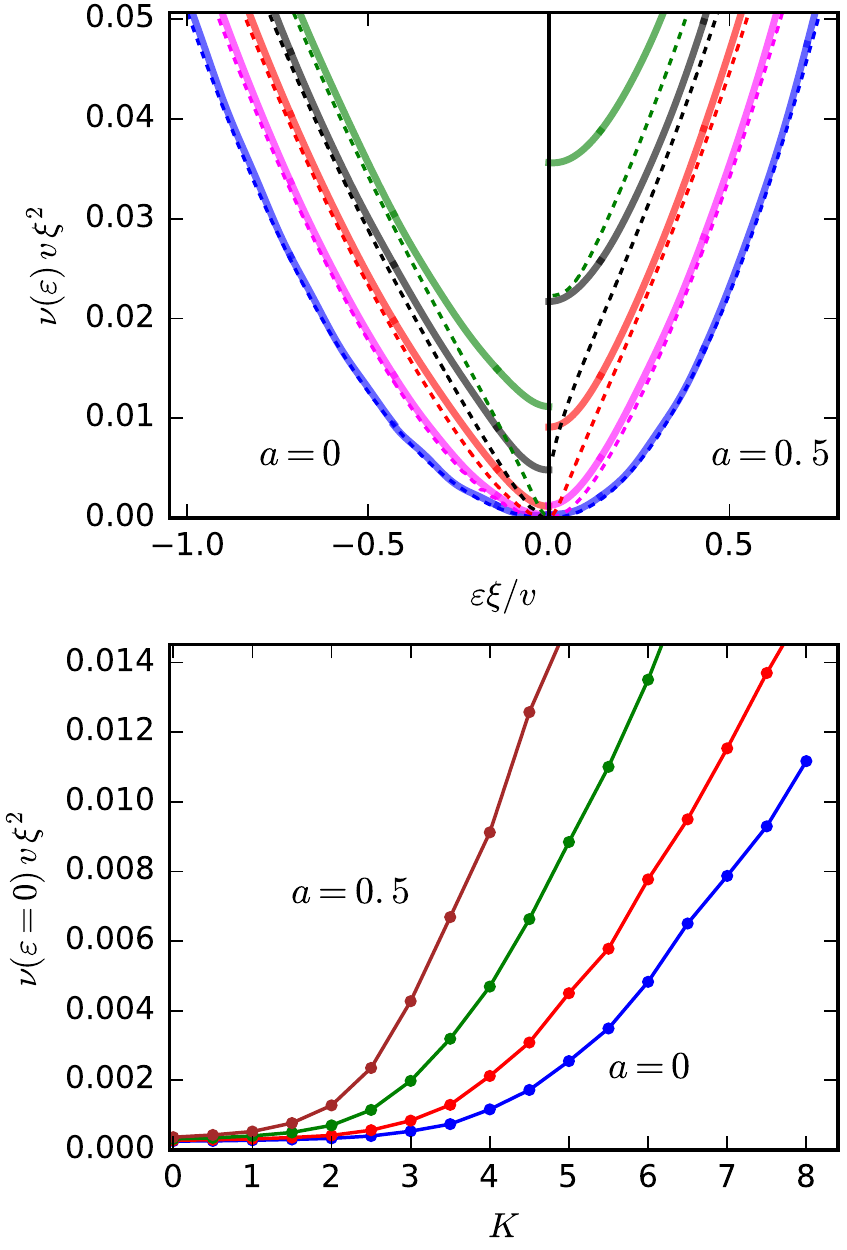}
\end{centering}
\caption{\label{fig:KPM} Top: Density of states $\nu(\varepsilon)$ as a function
of energy $\varepsilon$ without tilt (left, $a=0$) and with tilt (right, $a=0.5$),
as calculated by the kernel polynomial method for disorder strengths $K=0,2,4,6,8$ (bottom
to top). The dashed lines denote the SCBA results, Eq. \eqref{dos}. [For $K=0$ the SCBA coincides with the zero-disorder density of states of Eq. \eqref{dos-tilt}.]
Bottom: Density of states $\nu(0)$ at the nodal point
versus disorder strength for different values of the tilt, $a=0,\,0.25,\,0.4,\,0.5$ (right to left). The calculations were carried out for an average over $20$ realizations of the random potential, for a cubic lattice of size $L/b=200$, with $30$ random vectors for calculating the trace in the KPM and an expansion order of roughly 1000. The density of states is normalized to a single Weyl node. 
}
\end{figure}

In Fig.~\ref{fig:KPM}, top panel, we show the KPM results for the density of states $\nu(\varepsilon)=0$ without tilt (left panel, solid curves) and with dimensionless tilt $a = 0.5$ (right). We also show the predictions of the SCBA (dashed curves). For weak disorder and away from zero energy, the KPM
and SCBA results are in good agreement, but the comparison with the numerical results also shows that the SCBA overestimates the critical disorder strength $K_{\rm c}$ above which the zero-energy density of states becomes finite. The figure suggests, however, that this overestimation happens equally for zero tilt and for finite tilt, so that the trend predicted by the SCBA is indeed confirmed by the KPM. 

The bottom panel of Fig.\ \ref{fig:KPM} shows the density of states $\nu(0)$ at the nodal point as a function of the disorder strength for different values of the tilt parameter. Again, these results confirm the trend predicted by the SCBA, that larger tilt corresponds to a lower critical disorder strength. However, a precise determination of $K_{\mathrm{c}}$ is beyond
the capabilities of the KPM method.\cite{sbierski_quantum_2015}

\section{Quantum transport}
\label{sec:transport}

The disorder-induced phase transition is not only characterized by its effect on the density of states at the nodal point $\nu(\varepsilon)$, which is zero for $K < K_{\rm c}$ and finite for $K > K_{\rm c}$, the transition also strongly affects transport properties, such as the conductance or shot noise power. In this Section we show that the characteristic transport properties at the nodal point found for a Weyl dispersion without tilt persist for a tilted dispersion, and we provide more evidence in support of the result of the previous Sections, that the critical disorder strength is reduced in the presence of a tilted Weyl dispersion.

\begin{figure}[t]
    \begin{center}
        \includegraphics[width=0.5\textwidth]{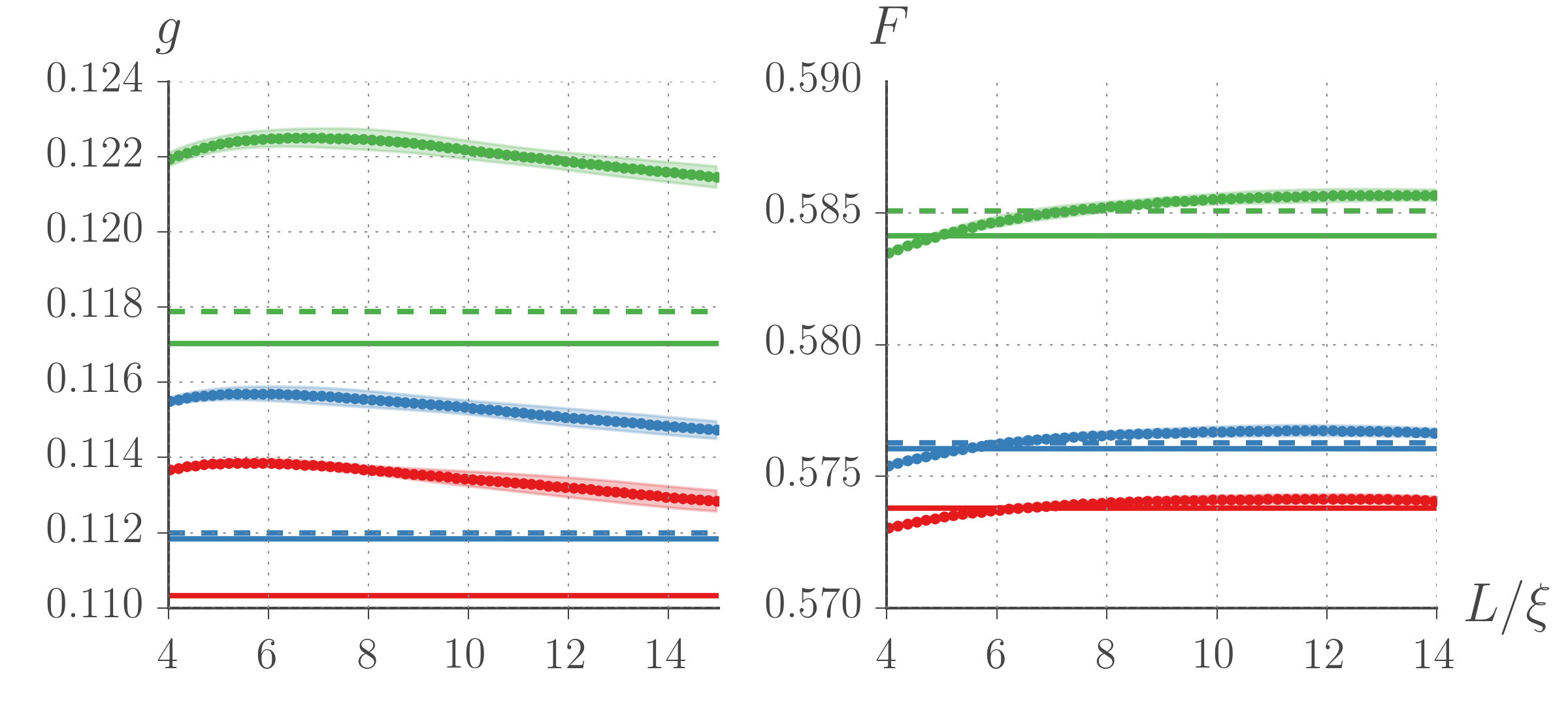}
    \end{center}
    \caption{(Color online) Dimensionless conductance $g$, rescaled for a system of size $L^3$ (top) and Fano factor $F$ for sub-critical disorder strength $K=1$ and for tilt parameter $a = 0$, $0.25$, and $0.5$ (bottom to top). The data points show results from numerical calculations, as described in the text. The solid and dashed line indicate the values expected without disorder and for a clean system with disorder-renormalized tilt $\tilde a$, respectively. We have not shown data for $L < 4$, which are dominated by finite-size effects. The numerical curves are based on an average over at least $10$ disorder realizations. The error bars show the residual statistical error. }
    \label{fig:smatrix-result}
\end{figure}
We calculate the transmission matrix $t$ of a finite sample of length $L$ in the transport direction (which we take to be the $x$-direction) and width $W$ in the transversal ($y$ and $z$) directions. The method is explained in detail in Ref.\ \onlinecite{sbierski_quantum_2014}. With a width $W$ the two-point correlation function (\ref{correlator-finite-size}) of the disorder potential is replaced by
\begin{align}
    \langle U(\vec{q}) U(\vec{q}')\rangle &= \frac{K \xi h^2 v^2}{W^2 L} e^{-\frac{q^2 \xi^2}{2}} \delta_{\vq,\vq'}.
\end{align}%
The conductance $G$ per Weyl node and the Fano factor $F$ are expressed in terms of $t$ as
\begin{align}
  G =& \frac{e^2}{h} \mbox{Tr}\, t t^{\dagger}, \\
  F =& 1 - \frac{\Tr (t t^{\dagger})^2}{\Tr t t^{\dagger}},
  \label{landauer}
\end{align}
where the trace is taken with respect to the transverse momenta. To approach bulk results as closely as possible, we choose the sample width $W \gg L$ and verify that the conductance is proportional to $W^2$ and is independent of the boundary conditions chosen in the transverse direction. We present results for the dimensionless conductance rescaled for a cubic sample of size $L^3$, $G = (e^2/h) (W^2/L^2) g$. The conductivity $\sigma$ is calculated from the standard relation $\sigma = (e^2/h) g/L$. The conductance $g$ and Fano factor $F$ without disorder, but with a tilted dispersion, were calculated in Ref.\ \onlinecite{trescher_quantum_2015}.

Figure \ref{fig:smatrix-result} shows the dimensionless conductance $g$ and the Fano factor versus system size $L$ for disorder strength $K=1$ and for the tilt parameter $a = 0$, $0.25$, and $0.5$. The disorder strength $K=1$ is sub-critical for all three values of $a$ considered. The figure also shows the conductance and Fano factor expected for a clean Weyl semimetal and for a clean Weyl semimetal with SCBA-renormalized tilt $\tilde a$. Especially the conductance data still show a considerable size dependence, even for the largest system sizes we could reach. Nevertheless, the data leave no doubt that the conductance remains bounded as a function of $L$, indicating that the conductivity $\sigma = (e^2/h) g/L$ is zero in the thermodynamic limit, which implies that the transport characteristics of the quasi-ballistic weak-disorder phase persist in the present of a tilted Weyl dispersion. Moreover, the finite-size data also show that the conductance increases with increasing tilt, consistent with the analysis of the clean limit in Ref.\ \onlinecite{trescher_quantum_2015} and with the expectation that tilt drives the system closer to the disorder-induced phase transition. However, the finite-size effects are too large to permit a more quantitative analysis. In particular, the finite-size effects are too large to quantitatively confirm or disprove the SCBA expectation for the conductance.

\begin{figure}[t]
    \includegraphics[width=0.47\textwidth]{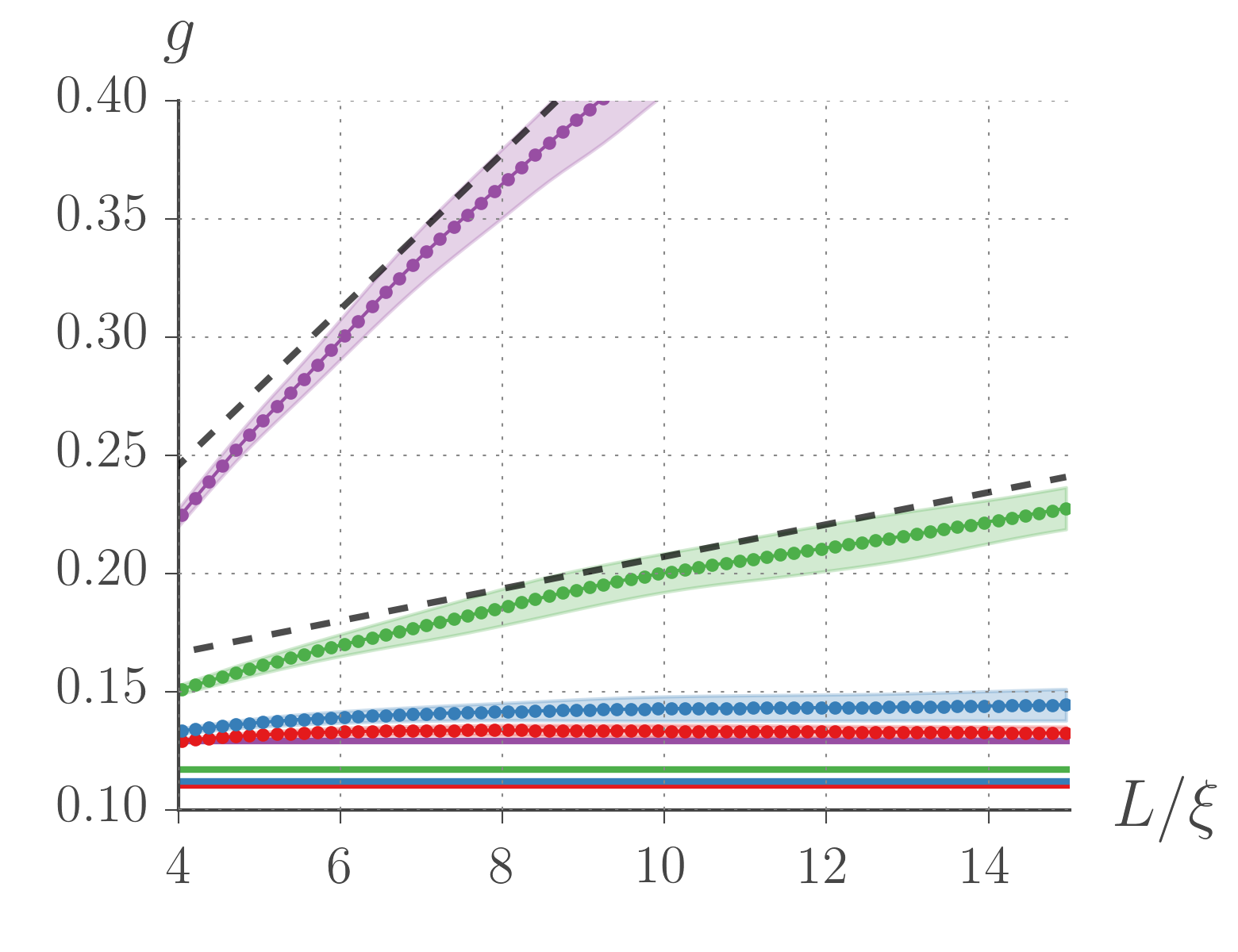}
    \caption{(Color online) Dimensionless conductance $g$ as a function of system size $L$ for disorder strength $K = 4$ and tilt parameters $a = 0$, $0.25$, $0.5$, and $0.75$ (bottom to top). The data points show results from numerical calculations after averaging over at least $10$ disorder realizations. The error bars show the residual statistical error. We have not shown data for $L < 4$, which are dominated by finite-size effects. The conductance is bounded for $a=0$ and $a=0.25$ (bottom two data sets), which is characteristic of the quasi-ballistic weak-disorder phase. For $a=0.5$ and $a=0.75$ the conductance is proportional to the system size $L$ for large $L$ (top two data sets), characteristic of the diffusive strong-disorder phase. The dashed lines indicate a linear increase with $L$ and are shown as a guide to the eye. }
    \label{fig:smatrix-result-medium}
\end{figure}

Figure \ref{fig:smatrix-result-medium} shows the dimensionless conductance $g$ versus system size for disorder strength $K = 4$, chosen such that the disorder is sub-critical without tilt, but above critical with large tilt. The progression of the curves shown in the figure corresponds to the horizontal arrow in Fig.\ \ref{fig:weyl-tilted} (although the value of the disorder strength $K$ is less than in Fig.\ \ref{fig:weyl-tilted}, reflecting the overestimation of the critical disorder strength in the SCBA). As in the previous figure, finite-size effects are considerable, nevertheless the asymptotic dependences characteristic of the weak-disorder phase ($g$ vs.\ $L$ bounded) and of the strong-disorder phase ($g \propto L$ for large $L$) are clearly visible for small and large tilt, respectively. Figure \ref{fig:smatrix-result-medium} thus illustrates how variation of the tilt can be used to scan through the disorder-induced phase transition between the quasi-ballistic weak-disorder phase and the diffusive strong-disorder phase. 
The Fano factor $F$ of the finite samples is subject to large fluctuations near the critical disorder strength and data is not shown.

\begin{figure}[t]
    \begin{center}
        \includegraphics[width=0.5\textwidth]{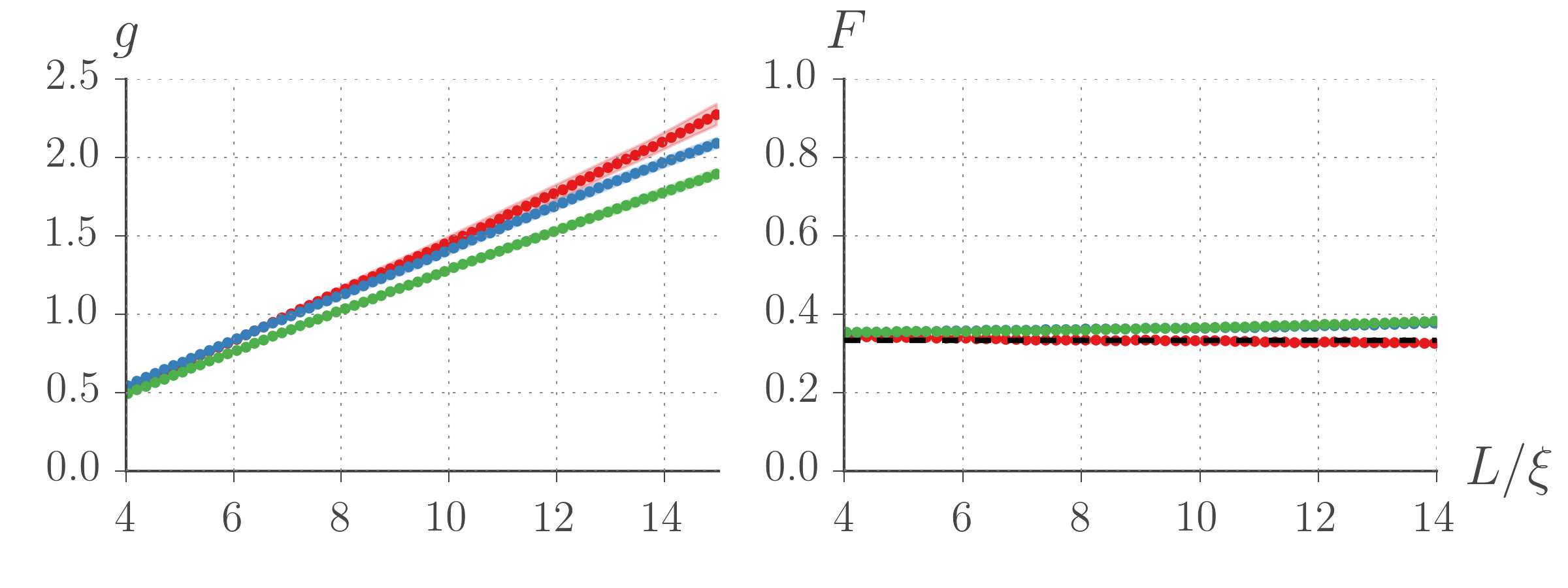}
    \end{center}
    \caption{(Color online) Dimensionless conductance $g$ and Fano factor $F$ as a function of system size $L$ for disorder strength $K = 30$ and tilt parameters $a = 0$, $0.25$, and $0.5$ (red, blue, green, i.e. top to bottom in the left panel). The data points show results from numerical calculations after averaging over at least $10$ disorder realizations. The error bars show the residual statistical error. Data for $L < 4$ are dominated by finite-size effects and are not shown. The dashed black line in the right panel indicates the Fano factor $F=1/3$ expected in the diffusive regime.}
    \label{fig:smatrix-result-strong}
\end{figure}

The case of very strong disorder is shown in Fig.\ \ref{fig:smatrix-result-strong}, which shows conductance and Fano factor for $K=30$, which is well above the critical value regardless of the tilt $a$. For all values of the tilt the conductance increases linearly with $L$, consistent with a finite conductivity $\sigma = (e^2/h) g/L$ in the thermodynamic limit. Up to small deviations (which we cannot explain, but which are ubiquitous in transport calculations) the Fano factor is at the value $F=1/3$ appropriate for the diffusive phase. In any case, our numerical calculations show that both conductance and Fano factor can be used to distinguish the weak-disorder and strong-disorder phases. In fact, in the presence of tilt the difference between the Fano factor for the weak-disorder and strong-disorder phases is even larger than without tilt, see Fig.\ \ref{fig:smatrix-result}.

\section{Discussion} 
\label{sec:discussion}

In this work we have studied how the combination of potential disorder and a tilted Weyl dispersion affects the zero-energy density of states and transport properties of a Weyl semimetal. Our main conclusions, supported via calculations of the density of states using the self-consistent Born approximation (SCBA), calculations of the density of states for a tight-binding model using the kernel polynomial method (KPM), and calculations of the conductance and Fano factor using the scattering approach, are that (1) the existence of a disorder-induced phase transition between a quasi-ballistic weak-disorder phase with zero density of states and finite conductance at the nodal point and a diffusive strong-disorder phase characterized by a finite conductivity and a finite density of states at the nodal point is unaffected by the presence of a tilt of the Weyl dispersion, and (2) the critical disorder strength for the transition decreases upon increasing the tilt. Importantly, the critical disorder strength approaches zero as the tilt approaches the critical tilt, so that in the presence of disorder the tilt-induced phase transition from a ``type-I'' Weyl semimetal (with a point-like Fermi surface) to a ``type-II'' Weyl semimetal (with finite particle and hole pockets at the nodal point) is always preceded by a disorder-induced transition to a diffusive phase with a finite density of states at the nodal point.

Another conclusion of our findings is that there is a family of weak-disorder pseudoballistic fixed points, all of them being effectively disorder-free, but distinguished by their effective tilt. The value of the tilt---{\em i.e.} which fixed point one is in---can be deduced from the Fano factor, which is a nontrivial function of the tilt parameter $a$.\cite{trescher_quantum_2015} (In principle, the dimensionless conductance $g$ also depends on the tilt but, unlike the Fano factor, $g$ also depends on a possible anisotropy of the Weyl dispersion, which the Fano factor $F$ does not.)

In principle tilted spectra can also occur in two-dimensional systems with a Dirac dispersion, such as graphene. In that context one expects that tilt, too, will quantitatively influence transport properties and the density of states. However, in two dimensions there is no disorder-induced phase transition between a quasi-ballistic and a diffusive regime, and the effect of tilt will be mainly quantitative, and not qualitative. Moreover, in graphene, certainly the two dimensional Dirac material studied the most, the Dirac points are at high symmetry points in the Brillouin Zone, at which anisotropies and tilts are prohibited by crystalline symmetries.

While our findings suggest that the tilt-induced transition between type-I and type-II Weyl semimetals may be obscured by the presence of disorder, they also suggest that tilt can be used as a parameter that drives the system through the disorder-induced transition between quasi-ballistic and diffusive phases. In general strain or pressure induces a change of the lattice geometry, which in turn influences the tilt. While the magnitude of the tilt is directly influenced by the hopping parameters of a lattice model,\cite{trescher_quantum_2015} the experimental feasibility depends on the specific material and the corresponding tilt's susceptibility to strain or pressure. Such limitations may make it necessary to start the tilt-induced phase transition already at a close-to-critical disorder strength, so that only a relatively minor change in the band structure is enough to drive the system through the quasi-ballistic-to-diffusive phase transition. (We note that pressure or strain may also result in a change of Fermi velocity, which would also influence $K_c$, and may be an additional factor that helps/obstructs the observation of the phase transition.)

\textit{ Acknowledgments.--}
This work is supported by 
DFG\textquoteright{}s Emmy Noether program (BE 5233/1-1) and CRC/Transregio 183 (Project A02) of the Deutsche Forschungsgemeinschaft, the Helmholtz VI ``New States of Matter and Their Excitations'', the Swedish research council and the Wallenberg Academy Fellows program of the KAW foundation.

\appendix

\section{}
\label{scba-appendix}

\begin{figure}
    \centering
    \includegraphics[width=0.99\linewidth]{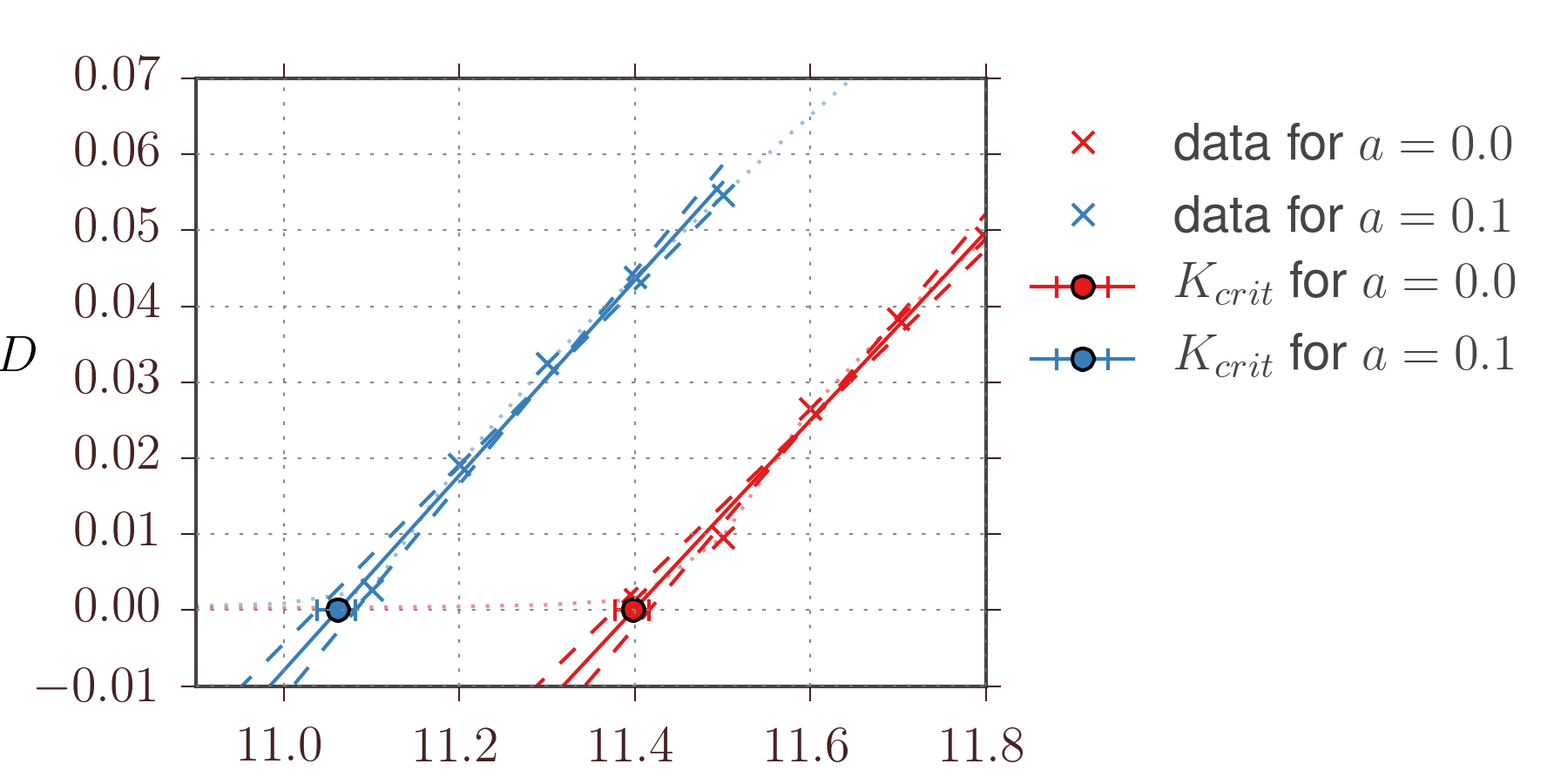}
    \caption{Linear fit to $D$ in the region where $D > D_{\mathrm{threshold}}$ to determine $K_{c}$.}
    \label{fig:determine-kcrit}
\end{figure}
In this appendix we provide further details on the calculation of the density of states using the self-consistent Born approximation (SCBA), see Section \ref{sec:scba}. Since there is rotational symmetry around the tilt direction (the $z$ axis), the parameter dependence of the self energy $\Sigma(\vk,\varepsilon)$ can be restricted, and $\Sigma(\vk,\varepsilon)$ can be parameterized as
\begin{align}
  \Sigma(\vk,\varepsilon) =&
  \Sigma_0(k,\theta,\varepsilon) +
  \Sigma_3(k,\theta,\varepsilon) \sigma_3 \nonumber\\ & \mbox{} +
  \Sigma_{\perp}(k,\theta,\varepsilon) (\sigma_1 \cos \varphi + \sigma_2 \sin \varphi),
\end{align}
where $\theta$ is the angle between $\vk$ and the $z$ axis and $\varphi$ is the azimuthal angle corresponding to $\vk$. Using the short-hand notations (with dependence on $k$, $\theta$, and $\varepsilon$ left implicit)
\begin{align}
    X &= \varepsilon - v a \cos(\theta) k - \Sigma_0(k, \theta, \varepsilon), \nonumber \\
    X_{\perp} &= v k \sin \theta
  + \Sigma_{\perp}(k,\theta,\varepsilon), \nonumber \\
    X_{3} &=
    v k \cos \theta + \Sigma_{z}(k, \theta, \varepsilon),
    \label{XY}
\end{align}
we can write the self-consistency condition for the Green function and self energy as

\begin{align}
    G =&\, \mbox{} \frac{1}{X^2 - X_{\perp}^2 - X_3^3} 
  \nonumber \\ & \mbox{} \times \left[
    X \sigma_0 + X_{\perp} (\sigma_1 \cos \varphi + \sigma_2 \sin \varphi)
    + X_3 \sigma_3 \right],
\end{align}
\begin{align}
    \Sigma_{\mu} =&\, \mbox{} 
    \frac{1}{(2\pi)^{3}} \int_{0}^{\infty} dk' \int_{0}^{\pi} d\theta' 
    \frac{X'_{\mu} \sin \theta'}{X'^2 - X_{\perp}'^2 - X_3'^2}
    B_{\mu}(\vk - \vk'),
    \label{XY-herleitung}
\end{align}
with $\mu = 0$, $\perp$, $3$ and
\begin{align}
  B_0(\vk-\vk') =&\, \mbox{} K \int_0^{2 \pi} d\varphi' e^{-|\vk-\vk'|^2 \xi^2/2}
  \nonumber  \\ =&\, \mbox{} 
  2 \pi K e^{-(k^2 + k'^2 - 2 k k' \cos(\theta) \cos(\theta')) \xi^2/2}
  \nonumber \\ &\, \mbox{} \times 
  I_0(2 k k' \xi^2 \sin \theta \sin \theta'), \\
  B_{\perp}(\vk-\vk') =& \mbox{} K \int_0^{2 \pi} d\varphi' e^{-|\vk-\vk'|^2 \xi^2/2}
  e^{-i \varphi}
  \nonumber \\ =&\, \mbox{} 
  2 \pi K e^{-(k^2 + k'^2 - 2 k k' \cos(\theta) \cos(\theta')) \xi^2/2} 
  \nonumber \\ &\, \mbox{} \times
  I_1(2 k k' \xi^2 \sin \theta \sin \theta'), \\
  B_3(\vk-\vk') =&\, \mbox{} B_0(\vk-\vk'),
\end{align}
where $I_0$ and $I_1$ are modified Bessel functions. Self-consistency equations for $X_0$, $X_{\perp}$, and $X_{3}$ then immediately follow from the definitions (\ref{XY}).

For the numerical computation we have to compute $G$ at small but finite imaginary values for the energy $\varepsilon$. Combined with the need to use a finite grid in momentum energy space to perform the integrations this leads to a slight rounding of the onset of a finite density of states upon entering the diffusive phase from the weak-disorder regime. This obstructs a direct determination of the critical disorder strength from the density-of-states calculations in the SCBA. We have estimated the precise location of the critical disorder strength from a linear fit of density of states for disorder strengths just above the critical disorder strength. An example of such a linear fit is shown in Fig.\ \ref{fig:determine-kcrit}. The fitting procedure leaves a small residual error for the value of $K_{\rm c}$, which is shown by the error bars in Fig.\ \ref{fig:weyl-tilted}.

\twocolumngrid

\end{document}